# Klipping's Ten Commandments

BELLAVE S. SHIVARAM

Department of Physics, University of Virginia, Charlottesville, VA., 22901, USA

and

USHA SINHA

Department of Physics, San Diego State University, San Diego, CA., 92182, USA

**ABSTRACT**

This is a short historical note about the visit of Prof. G. Klipping of Fritz Haber Institute, Germany, to the low temperature physics laboratory of Indian Institute of Technology, Madras (IIT/M). During his visit in 1979, Klipping delivered a series of lectures on cryogenic practices and low temperature physics in the physics department. The authors, both Master's degree students at that time, with a specialization in cryogenics, attended these lectures arranged by Prof. R. Srinivasan then head of the Low Temperature laboratory, IIT/M. In this non-technical note, the authors attempt to portray the research philosophy and attitudes of Prof. Klipping as captured through many wise remarks and snippets which formed an integral part of his lectures. In the 1970-80s Klipping made invaluable contributions to the development of cryogenics research in India.

It is common knowledge that the Indian Institute of Technology, Madras, (IIT/M) one of the first five premier post-independence era science and engineering educational institutions, was established with German assistance and partnership. Since its beginning in 1959 to about 1974, there were German professors who taught courses and oversaw laboratories at the Institute. The authors of this short article both entered IIT/M in 1977 as graduate students well after all the German faculty had departed. Nevertheless, a unique opportunity arose to interact with a well-known German professor. Prof. R. Srinivasan, then head and founder architect of the low temperature laboratory in the physics department of IIT/M arranged a two week visit by Prof. Gustav Klipping, of the Fritz Haber Institute, Germany, in the beginning of January 1979. Klipping and Srinivasan were collaborators, and the former was instrumental in assisting in the establishment of the low temperature physics laboratory at IIT/M[1]. This collaboration can be viewed in the broader context of post war German influence and outreach through joint efforts and conversations[2]. During the two-week visit Klipping delivered a series of lectures on cryogenics and low temperature physics. This is a brief account of the experiences and impressions the authors had while attending these lectures. These inspiring lectures along with the training they received in the low temperature laboratory played a pivotal role in the authors' and their fellow classmates' future pursuit of academic careers in this field (see caption to fig.1).

The Fritz Haber Institute in Germany has a rich history and celebrated its 100th year in 2011. Prof. G. Klipping who studied engineering at the same institute was picked by none other than Max von Laue, of X-ray diffraction fame, the then head of Fritz Haber to start a new liquid helium and liquid nitrogen facility[3]. Although Klipping came from an engineering background and started his career with this technical assignment he transitioned and eventually joined the Freie Universitat physics faculty in 1982. His published papers reflect his wide-ranging interests in both fundamental and applied aspects of cryogenics. His notable contributions include the development of the German Infra-Red space Laboratory (GIRL)[4], contributions to vacuum technology[5] and even to medicine[6], the latter jointly with Prof. Srinivasan. The two-week series of lectures he delivered at IIT/M had both aspects, critical engineering details of cryogenic methods and instruments as well as the relevant thermal and solid state physics. In addition to

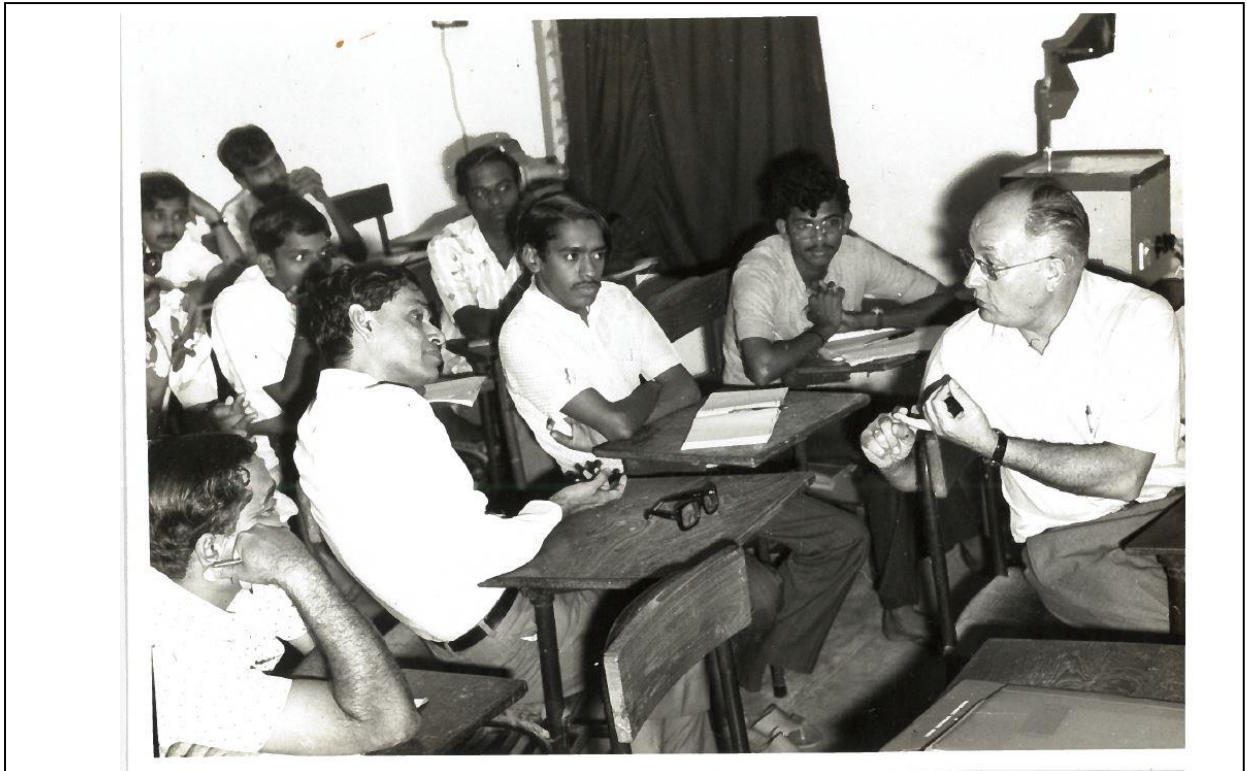

**Figure 1:** Photograph taken during the lecture given by Prof. Klipping. In the photograph are Prof. G. Klipping (extreme right); Prof. R. Srinivasan (second from left - spectacles down); Prof. T.P. Srinivasan (hands on chin); B. S. Shivaram (one of the authors, then M.Sc. student - seen with hand atop left ear); S. Ramakrishnan (left of Shivaram, also M.Sc. student and later Director of TIFR). The others are R. Ranganathan (Ph.D. scholar - behind R. Srinivasan) and V. Sankaranarayanan (Ph.D. scholar left of R. Ranganathan) who subsequently became professors. Also present in the lecture room but not in picture were Usha Sinha (author) and G. Rajaram, Professor of physics, Central University, Hyderabad.

### KLIPPING'S TEN COMMANDMENTS/COMMENTS:

You should be happy when you fail in your experiment, because that is going to make you think.

There are only few people who think...most people escape thinking.

There is no easy way to achieve higher standards...you must do evrything in steps...and..observe thoroughly step by step....

Progress is always not advantageous- you have to pay for every thing.

There is no escape- even the boss has to solder.

Thinking with the brain and working with the hand...you must must combine both these things. Then only new developments can be made.

Every physicist nowadays, especially younger people, want to do research with a lot of electronics,automation, on-line computing and all that. They come into the lab...like movie-stars..switch this on...that on...no understanding...no thinking in the brain....

You should transfer all your experience, all the graphs, to the lower parts of your brain. If you do something to your cryostat you must 'feel' what happens to the cryostat.

Even a dewar has a soul. In order to work with it you must try to understand it's soul.

Cryogenics- a must technologically for the future.

*I. Klipping*
6. Jan. 1979

**Figure 2:** Prof. Klipping's characteristic speaking style and his many wise sayings as captured by one of the authors (BSS).

the scientific and technical content of these lectures one could not escape coming out without being influenced by his infectious attitude about work, research style and priorities. The impact was compelling enough that one of the authors (BSS) felt it worthwhile to pen down some of his wise remarks and have them typed by the departmental secretary which Klipping was happy to sign (fig.2). These lectures were also memorable as Ingrid, his wife also a scientist and collaborator[7], and a German cryogenic technician, H. Walter, who had accompanied them would be present during the lectures and would frequently interject with entertaining and informative remarks.

The gist of Klipping's remarks - in addition to the typed up ten commandments - is perhaps relevant even today. Prof. R. Srinivasan who trained as a theoretical physicist at the Indian Institute of Science, Bangalore, transitioned to experimental physics in his IIT/M years. In addition to his talent for presenting ideas in the classroom very clearly and thoroughly he was very much a hands-on physicist. He could be seen late into the night working hand in hand with the then Ph.D. scholars in the low temperature laboratory. But we suspect that Klipping was sensitive to the broad difference in scientific cultures or styles between Germany and India. "There is no escape - even the boss has to solder" seems to be a remark which might have been the result of his perception that Indian physicists (of those times) hesitated to get their hands dirty. Klipping came across to us as a "hardliner" who insisted on "basics". "Even a dewar has a soul" implied that one has to spend enough time with the apparatus and come to terms with it. Although scientific instruments and pieces of apparatus in general are assembled following strict designs and protocols, it is possible that a particular instrument has its own quirks. Merely walking into a lab and flipping switches does not lead to scientific discoveries.

Prof. Klipping visited India at least one more time in later years as his influence on the development of cryogenics and low temperature physics in India had spread beyond IIT/M.  In 1982 he visited the Indian Institute of Science (IISc) in Bangalore when one of us (US) was a Ph.D. student.  This time he was accompanied by his daughter who also toured all the labs and interacted with the Indian researchers.  As we had come to expect from him, Prof. Klipping spent a significant amount of time in the cryogenics lab in his well-known hands-on fashion. At that time, the central area of the Physics Department at IISc was occupied by the machine shop, it was enroute to the cryogenics lab and hard to miss. Prof. Klipping's daughter observed that the female graduate students were standing close to the lathe machines while discussing with the technicians. She feared that the sari could result in accidents as the garment was flowing and there were no safety measures. Her observation about the need for heightened lab safety turned out to be prophetic when a small accident (luckily without injuries) occurred in the lab to one of the authors (US). This led to the last commandment that can be regarded as the daughter's eleventh commandment, ....'don't forget lab safety measures in the midst of a busy research routine'.

Lab safety is an issue that is finding increasing attention in the United States.  It is common practice in major US universities to have a student shop - a workshop with a full range of instruments such as lathes and milling machines useable by research students.  There have been such reports as a female student whose long hair was caught in a running lathe leading to loss of her life[8].  At the University of Virginia many decades ago access to such a student shop was very easy with the shop doors unlocked all times during working hours.  That laissez-faire situation has changed and access today is restricted to properly trained and authorized users only. We hope that Indian Universities with their developmental aspirations in the new growth environment pay due attention to lab safety particularly for female students.

**Acknowledgements**: We thank Prof. R. Srinivasan, Prof. G. Baskaran and Prof. Rolan Wittje for comments and clarifications. We also acknowledge the friendship and camaraderie of our classmates S. Ramakrishnan and G. Rajaram who went on to obtain their doctorates in low temperature physics and held important academic and research positions.

**REFERENCES**

[1] Roland Wittje, Engineering Education in Cold War Diplomacy: India, Germany, and the Establishment of IIT Madras, Ber. Wissenschaftsgesch, 2020, **43**, 560 – 580, DOI: 10.1002/bewi.202000014

[2] Hans Harder and Dhruv Raina (eds.), Disciplines and Movements: Conversations between India and the German Speaking World, Orient Blackswan, 2023.

[3] The Low Temperature Laboratory at the Fritz-Haber-Institut in Berlin, G. Klipping, I. Klipping, A. Sugawara and Y. Matsubara, Journal of Cryogenics and Superconductivity Society of Japan, 1978, **13**, 288-296,

[4] Dietrich Lemke, Gustav Klipping, Michael Grewing, Helge Trinks, Siegfried Drapatz, Klaus Proetel, "German Infra Red Laboratory (Girl)-Liquid Helium-Cooled Infrared Observatory For Spacelab", Proc. SPIE 0183, Space Optics II, 1979, 27 September; https://doi.org/10.1117/12.957394

[5] G. Klipping, "Relations Between Cryogenics and Vacuum Technology up to Now and in the Future", Japanese Journal of Applied Physics, 1974, **13**, 81, DOI 10.7567/JJAPS.2S1.81.

[6] G. Klipping, A. Krishna, U.Ruppert, R. Srinivasan and H. Walter, "Low temperature skin treatment", Cryogenics, 1981, **21**, Pages 181-185.

[7] Ingrid, Gustav Klipping, Mendelssohn Award Lecture: Our life with cryogenics, Chapter 1 - in Proceedings of the Twentieth International Cryogenic Engineering Conference

(ICEC20), Editor(s): Liang Zhang, Liangzhen Lin, Guobang Chen, Elsevier Science, 2005, Pages 1-9, ISBN 9780080445595, https://doi.org/10.1016/B978-008044559-5/50004-1.

[8] Lisa W. Foderaro, "Yale Student Killed as Hair Gets Caught in Lathe", New York Times, 2011, April 13.